\documentclass[12pt]{iopart}

\usepackage[utf8]{inputenc}
\usepackage[english]{babel}
\usepackage{csquotes}

\usepackage[T1]{fontenc} 
\usepackage{gensymb}
\usepackage{textcomp}
\usepackage{amssymb}
\usepackage{graphicx}
\usepackage{babel}
\usepackage[normalem]{ulem}
\usepackage[hidelinks]{hyperref}
\usepackage{appendix}
\usepackage{algorithm}
\usepackage{algorithmic}
\usepackage{cancel}
\usepackage{siunitx}
\usepackage{paralist}
\usepackage{placeins}
\usepackage{graphicx}

\usepackage{color}
\definecolor{blue}{rgb}{0,0,1}
\definecolor{black}{rgb}{0,0,0}

\definecolor{dgreen}{rgb}{0,0.5,0}

\definecolor{dred}{rgb}{0.5,0,0}

\definecolor{dyellow}{rgb}{0.75,0.75,0}

\begin{document}

\title{Universal properties of primary and secondary cosmic ray energy spectra}
\author{Marco Smolla$^{1, 2}$, Benjamin Sch\"afer$^3$, Harald Lesch$^2$ and Christian Beck$^3$}
\address{$^1$ Max-Planck-Institut für extraterrestrische Physik, Giessenbachstrasse, 85741 Garching, Germany}
\address{$^2$ Ludwig-Maximilians-Universität München, Universitäts-Sternwarte München, Scheinerstrasse 1, 81679 München, Germany}
\address{$^3$ School of Mathematical Sciences, Queen Mary University of London, London E1 4NS, United Kingdom}

\ead{\mailto{smolla@mpe.mpg.de}}
\ead{\mailto{b.schaefer@qmul.ac.uk}}
\ead{\mailto{c.beck@qmul.ac.uk}}

\begin{abstract}

Atomic nuclei appearing in cosmic rays are typically classified as primary or secondary. However, a better understanding of their origin and propagation properties is still necessary. We analyse the flux of primary (He, C, O) and secondary nuclei (Li, Be, B) detected with rigidity (momentum/charge) between $\SI{2}{\giga\volt}$ and $\SI{3}{\tera\volt}$ by the Alpha Magnetic Spectrometer (AMS) on the International Space Station. We show that $q$-exponential distribution functions, as motivated by generalized versions of statistical mechanics with temperature fluctuations, provide excellent fits for the measured flux of all nuclei considered. Primary and secondary fluxes reveal a universal dependence on kinetic energy per nucleon for which the underlying energy distribution functions are solely distinguished by their effective degrees of freedom. All given spectra are characterized by a universal mean temperature parameter $\sim \SI{200}{\mega\electronvolt}$ which agrees with the Hagedorn temperature. 
Our analysis suggests that QCD scattering processes together with nonequilibrium temperature fluctuations imprint universally onto the measured cosmic ray spectra, and produce a similar shape of energy spectra as high energy collider experiments on the Earth.

\end{abstract}


\submitto{\NJP}

\maketitle

\section{Introduction}
A fundamental challenge of current cosmic ray (CR) research is to understand the origin of highly energetic CRs, their
abundance in terms of different particle types,
and to identify the processes at work for acceleration and propagation. Collectiveley these processes determine the energy dependent flux of CRs, that is their energy spectra.
Because charged particles gyrate around the magnetic field lines of the interstellar medium (ISM), the directional information about the source is ultimately lost, leading to a roughly isotropic distribution observed here at Earth.
The atomic nuclei among the CRs are classified as primary CRs, usually thought to be expelled by supernovae explosions and accelerated in shock fronts of
supernova remnants, and secondary CRs, which result from particle collisions in the ISM. Here, we consider the flux of six different nuclei, namely the primaries He, C, O and the secondaries Li, Be, B as observed with the Alpha Magnetic Spectrometer (AMS) on the International Space Station \cite{Aguilar2017,Aguilar2018}.

It is commonly accepted that the major fraction of He, C, O can be classified as primary CRs whereas Li, Be, B are secondary CRs because their relative abundance exceeds the chemical composition of the ISM by a few orders of magnitude \cite{gaisser2016cosmic}. Some progress has been made in explaining CR acceleration (e.g. at supernova remnant shocks) \cite{Tatischeff2018} and propagation (e.g diffusion confinement) \cite{Strong2007} which allows to better investigate the specific processes responsible for the observed distributions. Nevertheless, considering the multitude of physical processes involved, our understanding remains incomplete and theoretical models accounting for the given nuclei spectra contain many unknown parameters and are currently under debate \cite{Gabici_2019}. 

As measured cosmic ray energy spectra decay in good
approximation with a power law over many orders of magnitudes, it is reasonable to apply a generalized statistical mechanics formalism (GSM) \cite{Wilk2010}
which generates power laws rather than exponential
distributions as the relevant effective canonical distributions.
Canonical Boltzmann-Gibbs (BG) statistics 
is only valid in an equilibrium context for systems
with short-range interactions, but it
can be generalized to a nonequilibrium context by introducing an entropic index $q$, where  $q>1$ accounts for heavy-tailed statistics and $q=1$ recovers BG statistics
\cite{tsallisbook,beckreview,Jizba2019}. The occurrence of the index $q$ can be naturally understood due to the fact that there 
are spatio-temporal temperature fluctuations in a general nonequilibrium situation, as addressed by the general concept of superstatistics, a by now standard statistical physics method \cite{beck2003superstatistics}.
Since the flux distribution as a function of energy in CRs evidently does not decay exponentially, it is reasonable not to use BG statistics but rather GSM, which
has been successfully applied to cosmic rays before in 
\cite{tsallis-borges, beck2004generalized,yalcin2018generalized} and also applied to particle collisions in LHC experiments \cite{beck2009superstatistics,Deppman2013,Wong2015}. Other applications of this superstatistical nonequilibrium approach are Lagrangian \cite{beck2007} and defect turbulence \cite{bodenschatz}, fluctuations in wind velocity and its persistence statistics \cite{rapisarda,weber2018wind}, fluctuations in the power grid frequency \cite{schafer2018non,Gorjao2020} and air pollution statistics \cite{williams2019superstatistical}. 

Here, we apply GSM and superstatistical methods to the observed CR flux of atomic nuclei to infer the physical parameters of the underlying energy distributions, which turn out to be nearly identical for all primaries and secondaries, respectively. The universal properties of the two CR types can be distinguished by a single parameter, the entropic index $q$, which we relate to the effective degrees of freedom of 
temperature fluctuations that are relevant in a GSM description.
The average temperature parameter
that fits all nuclei spectra turns out to be universal as well and is given by about $\SI{200}{\mega\electronvolt}$, coinciding with the Hagedorn
temperature. This suggests that QCD scattering processes play a dominant role in shaping the spectrum of observed cosmic
rays. The spectra are indeed similar to observed momentum spectra in high
energy proton-proton-collider experiments on the Earth, which are known to generate
$q$-exponential power laws \cite{Wong2015,beck2009superstatistics}. 

The paper is organized as follows: In section \ref{sec:Results} we demonstrate that cosmic ray nuclei spectra, as measured by AMS, are well described by $q$-exponential
distribution functions. We show that the
spectra exhibit data collapse if they are related to the kinetic energy per nucleon. In section \ref{sec:TemperatureFluctuations} we interpret the observed spectra in terms of temperature fluctuations occurring during the production process of the cosmic rays, based on 
$\chi^2$ superstatistics. We relate the power law spectral index to
the relevant degrees of freedom contributing
to the temperature fluctuations. Finally, a possible physical explanation 
of the universal properties of the observed spectra
is given in section \ref{sec:PhysicalInterpretation}.

\section{Results} \label{sec:Results}

\begin{figure*}
\centering
\includegraphics[width=\textwidth]{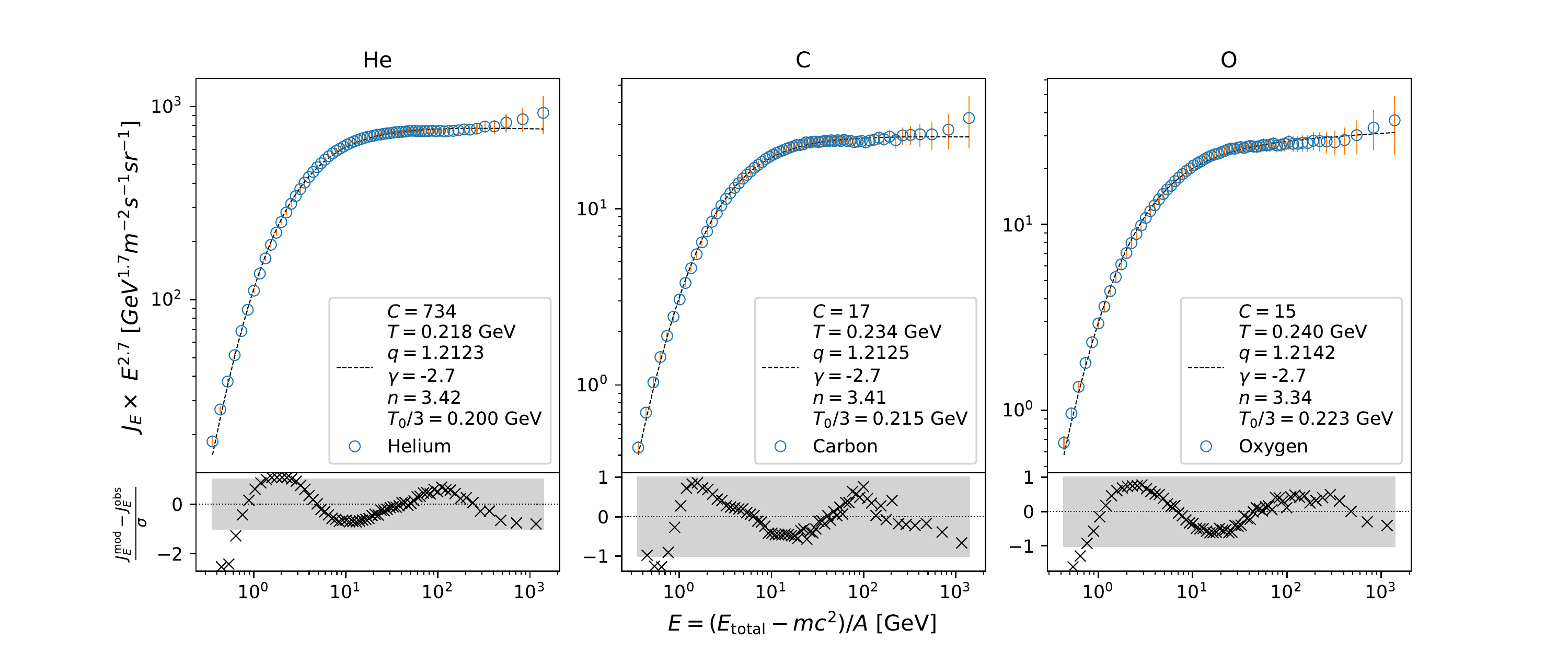}
\includegraphics[width=\textwidth]{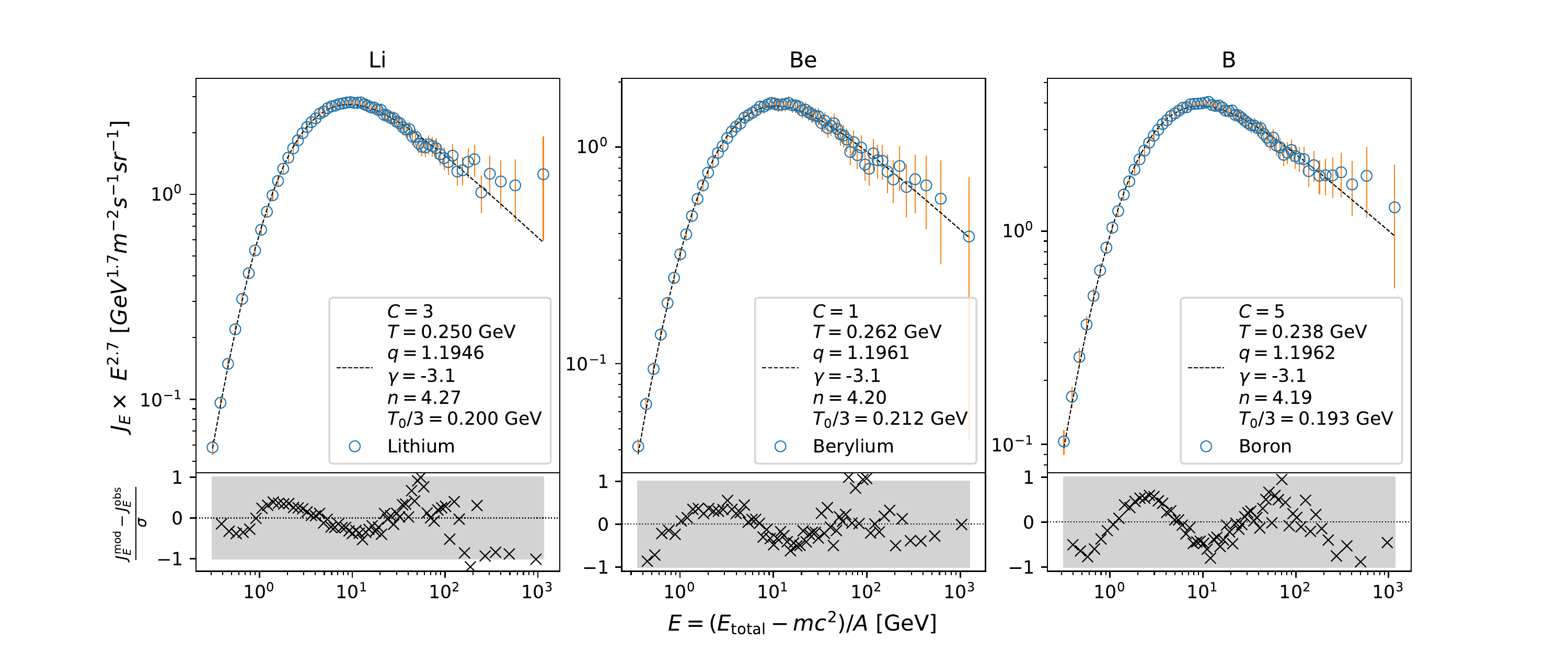}
\caption{The particle flux of each CR species was fitted with \eref{eq:FittingFunction} using three parameters $C, T, q$. The vertical axis in this log-log plot was multiplied with $E^{2.7}$ for better visibility. The fit's accuracy can be quantified by the deviation from modelled $J^{mod}$ to observed flux $J^{\text{obs}}$ weighted by the respective measurement error $\sigma$. Evidently, almost all data points fall within the uncertainty range of $\pm \sigma$ illustrated as grey shaded area. The mean temperature $T_0$ is defined by \eref{eq:mean_inverse_Temp}. The amplitude $C$ has dimensions $[C] = [\si{\meter\tothe{-2} \steradian\tothe{-1} \second\tothe{-1} \giga\electronvolt\tothe{-3}}]$.
}
\label{fig:SixPlots}
\end{figure*}

We investigate the cosmic ray flux, given as differential intensity with respect to kinetic energy per nucleon, defined as $E = (E_{total} - m)/ A$, with total energy $E_{\text{total}}=\sqrt{p^2 +m^2}$,
momentum $p= |\vec{p} |$, rest mass $m = Au$, mass number $A$, atomic mass unit $u=\SI{0.931}{\giga\electronvolt}$ and $[m] = [p] = [\si{\giga\electronvolt}]$ in  $c=1$ convention.

In order to infer physical parameters from the energy distribution fit to the observed cosmic ray flux, we employ an established GSM model \cite{beck2004generalized}, modified slightly by replacing the total energy by the kinetic energy per nucleon $E$. This choice of variable is common practice in CR literature because the kinetic energy per nucleon and the charge to mass ratio of a given particle provides the essential properties that decide how the particle’s trajectory will be modified by the presence of magnetic field lines. In the Appendix we rigorously derive the distribution function
\begin{eqnarray}
P_E (E)= C \rho (E) e_q^{-b E},
\label{eq:DistributionFunction}
\end{eqnarray}
which corresponds to the following differential intensity of flux
\begin{eqnarray}
J_E (E)= v(E) P_E,
\label{eq:DifferentialIntensity}
\end{eqnarray}
where $C$, $q>1$ and $T=b^{-1}>0$ are free parameters and the $q$-exponential is defined as $e_q^{x} \equiv \left(1 + (1-q)x\right)^{\frac{1}{1-q}}$ which implies $e_{q \to 1}^{x} = e^x$. $\rho (E)$ is a phase space factor which describes the density
of states, i.e. how many energy states can be taken on in a given range. For our fits we used $\rho (E) = p^2 \frac{\text{d}p}{\text{d}E}  = (E+u) \sqrt{E(E + 2u)}$ which leads to the flux derived from our superstatistical model
\begin{eqnarray}
    J_E^\text{mod} (E) = C E(E+2u) e_q(-b E),
    \label{eq:FittingFunction}
\end{eqnarray}
which we compare to the observed flux $J_E^\text{obs}$.

The entropic index $q$ determines the high-energy (i.e. the tail) behavior of the distribution since the $q$-exponential asymptotically approaches a power law
\begin{eqnarray}
\lim\limits_{E \to \infty} e_q^{-b E} \propto E^{\gamma},
\end{eqnarray}
with spectral index $\gamma = 2- 1/(q-1)$ for $q>1$. The parameter $T=b^{-1}$ represents a temperature in energy units that constrains the low-energy regime of maximum flux. Since our analysis focuses on the spectral \textit{shape} of the energy distribution we collect all global factors, which do not depend explicitly on the energy, in the amplitude $C$, which is merely a gauge for the absolute magnitude of the flux.

Fig. \ref{fig:SixPlots} illustrates that most data points are fitted by our model within a single standard deviation for all six nuclei. We determined the best fit by applying $\chi^2$ minimisation with $(J_E^{\text{mod}} - J_E^{\text{obs}})/\sigma$, meaning deviation of model from data weighted by the respective measurement uncertainty, where the standard deviation $\sigma$ is the sum of measurement errors for a specific energy bin. For most of the data the error is of the order of a few percent whereas the uncertainty tends to increase with energy up to the largest uncertainty of $89$ percent associated with the Beryllium flux measured in the highest energy bin.

\begin{figure}
\includegraphics[width=\textwidth]{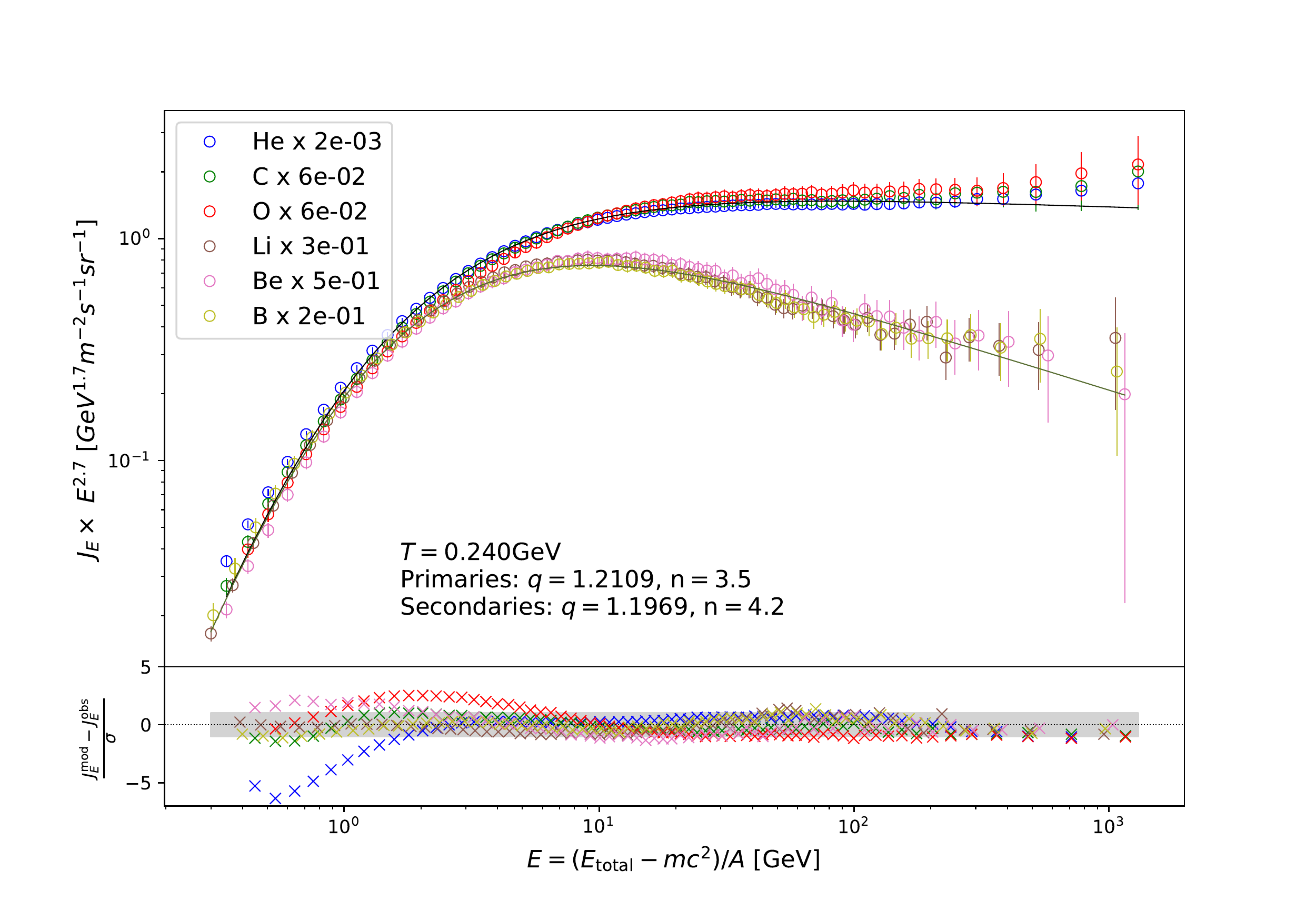}
\caption{Each particle flux was rescaled with a suitable factor such that the data points (roughly) collapse to a single line at the low energy end and the universal properties of primary and secondary cosmic ray nuclei spectra become visible. For larger energies the spectrum splits into primaries and secondaries which can be distinguished by a single parameter, the entropic index $q$ which can be interpreted by the underlying effective degrees of freedom.
}
\label{fig:Universality}
\end{figure}

Fig. \ref{fig:Universality} reveals the universal properties of the primary (He, C, O) and secondary (Li, Be, B) cosmic ray fluxes when rescaling each nuclei flux with a suitable global factor such that all data points collapse 
to a single line in the low energy range. 
Fixing the global amplitude parameter to $C=1$ and $T= \SI{0.240}{\giga \electronvolt}$, which is the average value for the temperatures inferred from the individual best fits in Fig.~\ref{fig:SixPlots}, allows us to do a best fit with $q$ as the only free parameter for the collapsed data of primaries and secondaries. This yields $q_{\text{prim}}= 1.2109$ ($n = 3.5$) and $q_{\text{sec}}= 1.1969$ ($n = 4.2$), where $n$ can be interpreted as degrees of freedom of temperature fluctuations as outlined below.

\section{Interpretation in terms of temperature fluctuations} \label{sec:TemperatureFluctuations}
We consider the observed cosmic ray spectra to be the result of many
different high energy scattering processes, each having a different local
temperature $\beta^{-1}$ in the local scattering
volume. This idea was previously worked out in detail for collider
experiments using LHC data, e.g. in \cite{beck2009superstatistics}. There are strong fluctuations
of temperature in each scattering event, which can be described by superstatistics \cite{beck2003superstatistics}, a standard method in the theory of complex systems. For cosmic rays, we need to generate asymptotic power laws and this can be achieved by so-called $\chi^2$ superstatistics.
As is generally known (see, e.g. \cite{beck2004generalized}) the probability density function for a fluctuating $\beta$ of the form
\begin{equation}
\beta = \sum_{i=1}^n X_i^2  \label{eq:sumbeta}
\end{equation} 
with independent and identically distributed Gaussian random variables $X_i$ is a $\chi^2$ distribution given by  
\begin{eqnarray}
\label{eq:ChiSquare}
  g(\beta) = \frac{1}{\Gamma (n/2)} \left( \frac{n}{2 \beta_0}\right)^{\frac{n}{2}} \beta^{\frac{n}{2} -1} \exp{\left(- \frac{n \beta}{2 \beta_0}\right)}.
\end{eqnarray}

It is well-known in the formalism of superstatistics that  superimposing various subsystems with different temperature weighted with $g(\beta)$ leads to $q$-exponential statistics.
For each scattering event
we apply ordinary statistical mechanics locally,
i.e. the conditional probability density of a kinetic
energy state $E$ in a given scattering event for a given
temperature is
\begin{equation}
    p_E(E|\beta)=\frac{1}{Z(\beta)} \rho (E) e^{-\beta E}.
    \label{eq:MarginalProb}
\end{equation}
In order to normalize our conditional distribution function we need to integrate over all possible energy states, obtaining the normalization constant
$Z(\beta) = \int_0^\infty \rho(E) e^{-\beta E}dE$. The marginal distribution $P_E(E)$ (the unconditioned distribution of energies) can be computed by integrating the conditional distribution $p_E(E|\beta)$ over all inverse temperatures $\beta$ weighted with $g(\beta)$. In the relativistic limit (neglecting mass terms) this yields 
\begin{eqnarray}
 P_E(E) &= \int_0^\infty g(\beta)
 p_E(E|\beta)\mathrm{d}\beta
 &\sim \rho(E) e_q^{-b E}
\end{eqnarray}
with $b = \beta_0 / (4-3q)$ and mean inverse temperature
\begin{equation}
    \beta_0 = \int_0^{\infty} \beta g(\beta) \text{d}\beta =: \frac{1}{T_0}.
    \label{eq:mean_inverse_Temp}
\end{equation} The effective degrees of freedom $n$ are related to the entropic index $q$ via
\begin{eqnarray}
\label{eq:dof}
    n = \frac{2}{q-1} - 6.
\end{eqnarray}

Considering the physical meaning of the random variable $X_i$ defined in equation (5) there are different possible interpretations, see \cite{beck2004generalized, beck2007, Beck2001, gravanis2020physical} for details. The main conclusions in our given analysis are independent of the particular interpretation chosen but it is worth emphasizing that it is physically plausible to understand $X_i$ as a measure for the fluctuating effective energy dissipation, with $n=3$ representing the three spatial degrees of freedom as minimum value, which is increased to $n=4$ when including variations in time.

In the Appendix we provide a more detailed derivation of the above results, which show that $q$-exponential statistics follows naturally from summing up ordinary Boltzmann distributions with $\chi^2$-distributed inverse temperatures.

\section{Physical interpretation and possible reason for universality} \label{sec:PhysicalInterpretation}

For the temperature parameter
$T_0=\beta_0^{-1}$, defined in (\ref{eq:mean_inverse_Temp}),
we get the value $T_0 \sim$ 600 MeV for each of the six CR species in our fits. Hence, the average effective temperature per quark is of the order $T_0/3 \sim \SI{200}{\mega\electronvolt}$, i.e. we recover approximately the observed value of the Hagedorn temperature which is roughly known to be in the range \SIrange[range-units = single]{140}{200}{\mega \electronvolt} \cite{hagedorn1965statistical, Rafelski2016, Broniowski2004} and represents a universal critical temperature for the quark gluon plasma and for high energy QCD scattering processes. Remarkably, the fitted value of $T_0/3$ in the fits is observed to be the same for all six nuclei, i.e. for both primary and secondary cosmic rays within a range of about one tenth of its absolute value.\\
Let us provide some arguments on why we consider $T_0/3$ as a relevant temperature parameter. In general, there are two alternative formulations of superstatistics, defined as type A and type B in \cite{beck2003superstatistics}, which yield the same form of distribution functions but differ in their definitions for $T_0$ because type A uses the unnormalized whereas type B uses the normalized Boltzmann factor for deriving the generalized canonical distribution, as we present in detail for type B in the appendix. We consider type B superstatistics as physically more plausible because we understand the generalized canonical distribution as originating from a superposition of many cosmic ray ensembles associated with a normalized canonical distribution respectively. Since the Hagedorn temperature is associated with the kinetic energy of particles interacting via the strong nuclear force, we divide the average temperature $T_0$ by the number of quarks, namely three for atomic nuclei. For a quark-gluon plasma one can either define a temperature for single quarks, or - after hadronization - for mesonic or baryonic states. At the critical Hagedorn phase transition point, where both states exist, this is mainly a question of definition \cite{Beck2000}.

The emergence of the Hagedorn temperature (at least as anorder of magnitude) in our fits suggests that cosmic ray energy spectra might originate from high energy scattering processes taking place at the Hagedorn temperature $T_H$. Very young neutron stars, initially formed in a supernova explosion, indeed have a temperature $\sim \SI{E12}{\kelvin} \sim \SI{100}{\mega\electronvolt}$ of comparable order of magnitude as the Hagedorn temperature \cite{lattimer}.
As the Hagedorn temperature is universal, so is the
average kinetic energy per quark of the cosmic rays nuclei, assuming they are produced in a Hagedorn fireball, either during the original supernova explosion, or later in collision processes of highly energetic CR particles with the ISM.

Our observation that the kinetic energy per nucleon (or per quark) yields 
universal behavior of the spectra is indeed pointing towards QCD processes as the dominant contribution that shapes the spectra
(see also \cite{yalcin2018generalized}): Were
there mainly electromagnetic processes underlying the
spectra, one would expect invariance under
rescaling with $Z$, but we observe invariance (universality)
under rescaling with $A$.
At the LHC one observes similar $q$-exponentials for the measured
transverse momentum spectra, as generated
by QCD scattering processes, with a temperature parameter $b^{-1}$ that
is of similar order of magnitude ($\SI{150}{\mega\electronvolt}$) as in our fits for the
cosmic rays, see table IV in \cite{Wong2015}. That paper
also shows that hard parton QCD scattering leads to power law spectral behavior.

Note that while the entire energy spectrum in figure 1 is well fitted by a $q$-exponential, the residuals tend to oscillate. A similar oscillatory behavior of the residuals (logarithmically depending on the energy) has been observed in the transverse momentum distribution for high energy $pp$ collision experiments at the LHC \cite{Wilk2017, Wong2015}. The similarity of these log-periodic oscillations for our cosmic ray data and for collider experiments on the Earth is indeed striking, and once again supports our point that both phenomena could have similar roots based on high energy scattering processes.

After having analysed the average temperature, let us now concentrate on the \textit{fluctuations} of temperature
in the individual  scattering events, described by the
parameter $n$, which determines the entropic index $q$ and thereby the tail behavior. One readily notices
that in our GSM model the marginal distributions $P_E(E)$
decay asymptotically as
\begin{equation}
P_E(E) \sim E^{-1-\frac{n}{2}}.  \label{10}
\end{equation}

In order to calculate the expectation of the fluctuating energy $E$,
\begin{equation}
\langle E \rangle =\int_0^\infty E P_E(E) \text{d}E,
\end{equation}
one notices that the integrand decays as $EP(E) \sim E^{-n/2}$.
Thus the expectation value is only well defined if $n>2$.

In the absence of further effects, like an energy dependent cross section, we could explain $n$ exclusively by the underlying statistics and thus associate $n$ with the number of Gaussian random variables contributing to the fluctuating $\beta$ in equation \eref{eq:sumbeta}. Since for cosmic ray propagation energy dependent processes affect the spectral shape the derived value for $n$ will represent both the statistical properties and the energy dependent processes. For this reason effective non-integer values for $n$ are possible. Because the above argument about the existence of the expectation value should apply more generally, and thus even in the absence of additional spectral modifications, we conclude that n = 3 is the minimum value for the degrees of freedom.

A similar argument applies if
one looks at the existence of the mean of the temperature
$\beta^{-1}$ as formed with the probability density $g(\beta)$
\begin{equation}
    \langle \beta^{-1} \rangle =\int_0^\infty \beta^{-1} g(\beta) \text{d} \beta.
    \label{eq:MeanTemperature}
\end{equation}
The above mean
only exists for $n>2$, since the integrand behaves as
$\beta^{\frac{n}{2}-2}$ for $\beta \to 0$.
We are thus naturally led to the minimum value $n=3$ as the strongest
fluctuation state of the Hagedorn fireball for which
a mean energy $\langle E \rangle$ and a mean temperature $\langle \beta^{-1} \rangle$ is well-defined. 

For secondary cosmic rays, there is an additional degree of freedom
as an additional collision process at a later time is needed to produce secondary cosmic rays. 
Thus it is plausible that for secondary cosmic rays $n$ is larger than the
minimum value $n=3$. 
The next higher value of $n$, which can be regarded as an excited state of temperature fluctuations, $n=4$, corresponds to $q=1.2000$. Indeed, based on our fits (see Fig. \ref{fig:SixPlots}), secondary cosmic rays are well approximated by this $q$ and therefore $n=4$. 

In the experimental data detected by AMS, it is to be expected that we will not observe the exact values of $q=1.2222$ and $q=1.2000$ since the spectra are modified by diffusion processes in the galaxy, by energy dependent escape processes from the shock front of the accelerating supernova remnant, and by radiative losses from acceleration.
All these effects can alter the spectrum and lead to small changes in the optimum fitting
parameters $q$ and $T$. We think this is the reason why
the best fits of the observed spectra correspond to $n=3.5$ rather than $n=3$, and $n=4.2$
rather than $n=4$, equivalent to minor negative corrections for the
spectral index $\gamma$ of the order $\Delta \gamma  \approx -0.1$. Also, the effective temperature $T$ may be increased
by diffusion processes in the galaxy, which will 
broaden the distributions. However, it seems these effects are only small perturbations that slightly modify the 
universal parameters set by the QCD scattering processes.

While the connection of QCD and generalized statistical mechanics was emphasized in \cite{Wong2015}, the model that we implement in our paper is mainly based on a nonequilibrium statistical mechanics approach, as originally introduced in \cite{beck2004generalized}. This  approach is based on temperature fluctuations in each (small) interaction volume, where the scattering event takes place. These local temperature fluctuations are at the root of the observed $q$-exponentials and the associated temperature scale turns out to coincide approximately with the Hagedorn temperature known from QCD. Other authors \cite{Deppman2016} have emphasized the fractal and hierarchical structure of scattering events and hadronization cascades, or the complexity of long-range interactions in the hadronization process \cite{Bediaga2000, Beck2000} coming to similar conclusions.

\section{Discussion on relevance of solar wind modulation}
\label{sec:SolarWind}

\begin{figure*}
\includegraphics[width=\textwidth]{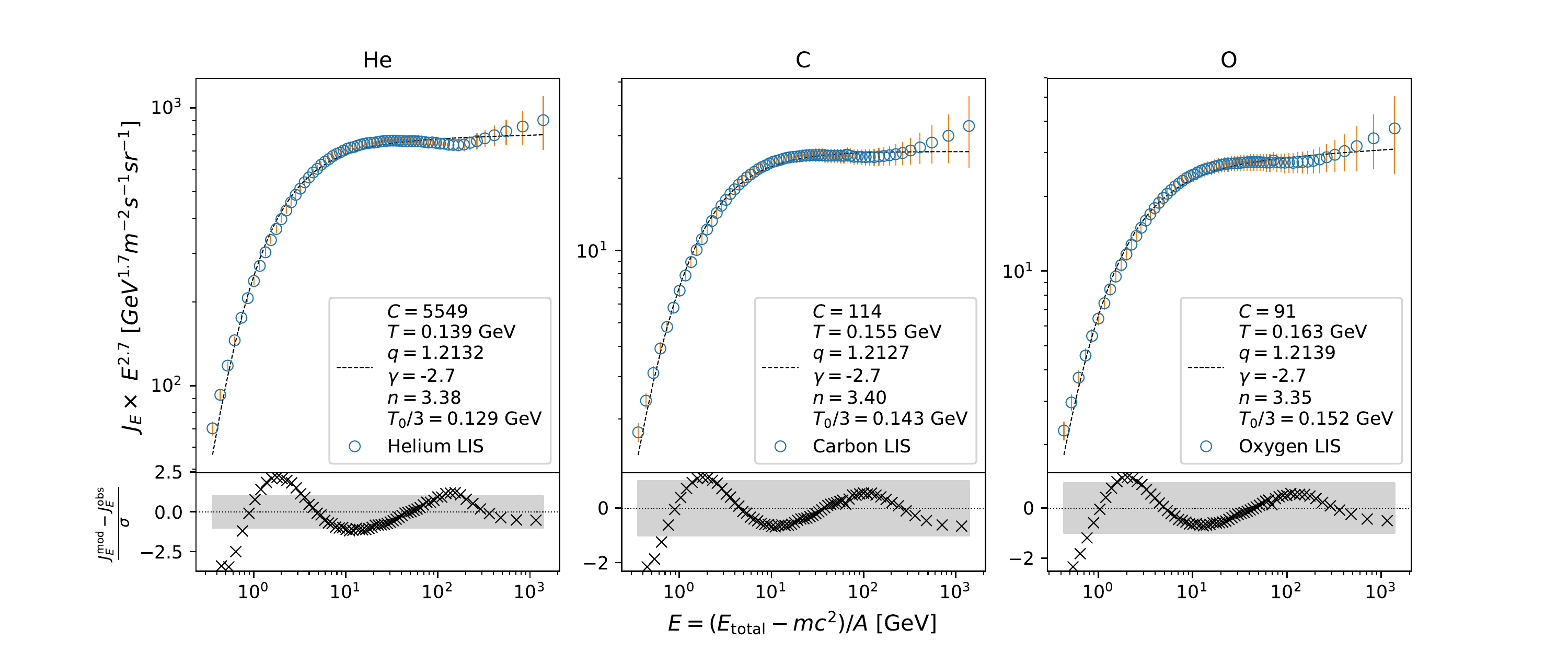}
\includegraphics[width=\textwidth]{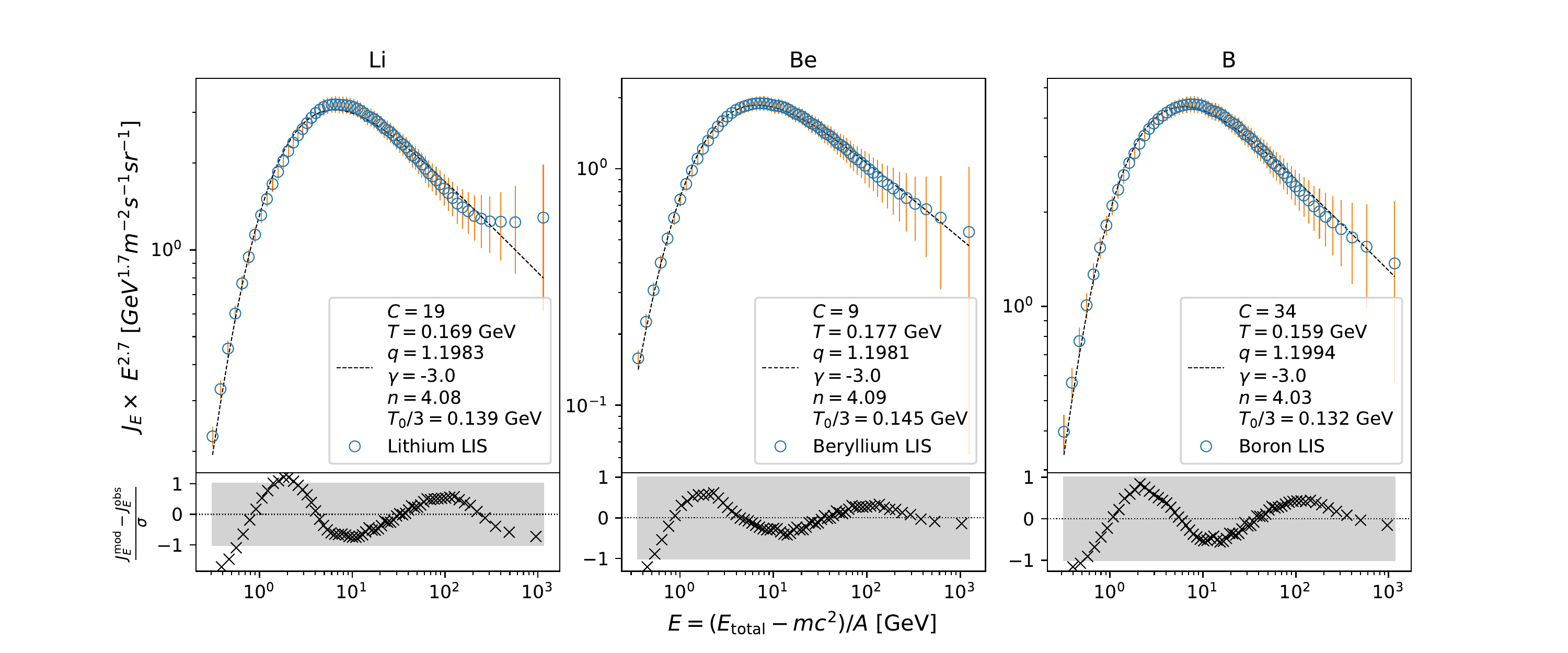}
\caption{Superstatistical results are robust when correcting for heliospheric impact.  Using cosmic ray propagation models HelMod and Galprop allows \cite{Boschini2017, Boschini_2018, Boschini2020} to estimate flux outside the heliosphere, that is unmodulated by solar wind representing the local interstellar spectra, in short LIS. Here, we use the data published in \cite{Boschini2017, Boschini_2018, Boschini2020} which we investigate for the given AMS energy bins. We apply $J_E^\text{mod} (E) = C E(E+2u) e_q(-b E)$ with $e_q^{x} \equiv \left(1 + (1-q)x\right)^{\frac{1}{1-q}}$ in order to derive the best fit global amplitude $C$, temperature $T$ and entropic index $q$. From the best fit parameter $T$ we derive the average temperature per quark as $T_0/3$. The entropic index q can be translated into effective degrees of freedom $n$ and into the spectral index $\gamma$ representing the asymptotic power law behaviour $\lim\limits_{E \to \infty} e_q^{-b E} \propto E^{\gamma}$.}
\label{fig:LIS-SixPlots}
\end{figure*}

The AMS measurements were taken at about $\SI{400}{\kilo \meter}$ above Earth's surface and are thus subject to solar wind modulation which yields a suppressed flux compared to outside the heliosphere, in particular for charged particles with kinetic energies per nucleon below $\lesssim \SI{10}{\giga \electronvolt}$ \cite{Moraal2014}. Thus for our given AMS data with kinetic energies per nucleon in the range of $\SI{0.4}{\giga \electronvolt} \lesssim E \lesssim \SI{1.2}{\tera \electronvolt}$ we would like to quantify the effect of solar wind modulation on our given spectra. Using cosmic ray propagation models allows to infer the unmodulated flux before cosmic rays are entering the heliosphere, that is the local interstellar flux.\\
This was recently done by \cite{Boschini2017, Boschini_2018, Boschini2020} who combined the two cosmic ray propagation models HelMod and Galprop and published the calculated flux for all our given atomic nuclei and for the entire energy range covered by AMS. We use their data and interpolate it to match the AMS energy bins definition. We find a maximum deviation of local interstellar flux (LIS) to flux inside heliosphere (AMS) for the lowest energetic particles with $LIS/AMS \lesssim 4$ for $E=\SI{0.4}{\giga \electronvolt}$. The two spectra converge for larger energies quickly and we find $LIS/AMS \lesssim 1.2$ for $E=\SI{10}{\giga \electronvolt}$ such that in fact only the lowest energy range of our spectra is significantly affected. Since the propagation model provides the flux without giving any uncertainty, we assign each estimated flux the same relative error as in the AMS data set. This makes the comparison between the flux inside and outside the heliosphere consistent and allows to put appropriate weight on measurements with smaller uncertainties for our least-square optimization.\\
Analogously to the steps performed for the given AMS data, we apply our generalized statistical mechanics methodology to the LIS data and present the resulting fits and parameters in figure \ref{fig:LIS-SixPlots}. The average temperatures $T_0/3$ for the different nuclei are about $50$ to $\SI{80}{\mega \electronvolt}$ lower than for the AMS data, namely in the range \SIrange[range-units = single]{129}{152}{\mega  \electronvolt}. Still, these temperatures are all about the scale of the Hagedorn temperature and in fact coincide with the temperature range \SIrange[range-units = single]{130}{160}{\mega \electronvolt} inferred by GSM methods applied to LHC experiments found by \cite{Wong2015, Wilk2017}. The effective degrees of freedom remain approximately the same. Since the reliability of our methodology ultimately depends on having a large energy range measured for all the different nuclei, the AMS data is the best currently available experimental data set. In contrast, measurements acquired by Voyager outside the heliosphere only cover energies from about $\SI{3}{\mega \electronvolt}$ to a few hundred MeV \cite{Cummings2016, Stone2019}. Hence, we apply our analysis to the large range of AMS-measured data and estimate the modulation by the solar wind, rather than using theoretically derived data for unmodulated spectra.

\section{Conclusion}

We provide excellent
fits for the measured AMS spectra of primary (He, C, O) and secondary cosmic rays (Li, Be, B)
using a simple superstatistical model.
The observed $q$-exponential spectra are interpreted in terms of temperature fluctuations occuring in the Hagedorn
fireball during the production process of cosmic rays
in their individual scattering events.
We provide evidence that the observed spectra of CR nuclei share universal properties: The spectra collapse if the kinetic energy per nucleon is taken as the relevant variable.
Primary and secondary CRs can be uniquely distinguished by their respective entropic index $q$, corresponding to different degrees of freedom associated with the temperature fluctuations. They share the same average temperature
parameter, whose order of magnitude coincides with the Hagedorn temperature.

\ack{Acknowledgements}
This project has received funding from the European Union’s Horizon 2020 research and innovation programme under the Marie Sklodowska--Curie grant agreement No 840825. In addition, we acknowledge support from the DFG Cluster of Excellence 'ORIGINS'.

\appendix
\setcounter{section}{0}

\section{Deriving the superstatistical distribution function}
We derive the distribution function \eref{eq:DistributionFunction}, that is $P_E = C \rho (E) e_q^{-b E}$, using the framework of superstatistics by which we can interpret the best fit parameters with a temperature $T=b^{-1}$ and  effective degrees of freedom $n$.

Superstatistics \cite{beck2003superstatistics} is a generalization of Boltzmann statistics in the sense that the distribution function can be derived by integrating the conditional probability distribution $p_E(E|\beta)= \rho (E) e^{-\beta E} / Z(\beta)$ for all given values of inverse temperature $\beta$. The normalization is calculated by summing over all possible energy states, yielding $Z(\beta) =  \int_{0}^{\infty} \rho(E) e^{-\beta E} \text{d}E$. In agreement with \cite{beck2004generalized} we apply the ultra-relativistic approximation for the density of states $\rho(E) \sim E^2$ in order to calculate $Z(\beta) \sim \beta^{-3}$. Given the $\chi^2$-distributed $\beta$, defined by \eref{eq:ChiSquare}, we calculate the generalized canonical distribution as follows:
\begin{eqnarray}
    P_E(E)&= \int_0^\infty g(\beta) p_E(E|\beta)\text{d}\beta \\
    &\sim \left( \frac{n}{2 \beta_0}\right)^{\frac{n}{2}} \rho(E)  \int_0^\infty  \beta^{\frac{n}{2} +2} e^{-\beta \left( E + \frac{n}{2\beta_0}\right)} \text{d}\beta \\
    &\sim \rho(E)  \left( \frac{n}{2 \beta_0}\right)^{\frac{n}{2}} \left( E + \frac{n}{2 \beta_0} \right)^{-3 - \frac{n}{2}}.
\end{eqnarray}
Introducing $q = 1 + 2 / (n+6)$ (equivalent to $n/2 = 1/(q-1) - 3$) and $b = \beta_0 / (4-3q)$, allows us to express the result as:
\begin{eqnarray}
    P_E(E) &\sim \rho(E) \left( E + \frac{n}{2 \beta_0} \right)^{-3}
    \left( \frac{\frac{n}{2 \beta_0}}{E + \frac{n}{2 \beta_0} } \right)^{\frac{1}{q-1}-3}  \\
    &\sim \rho(E)  \left( \frac{n}{2 \beta_0} \right)^{-3} 
    \left( \frac{1}{1 + E \frac{2 \beta_0}{n} } \right)^{\frac{1}{q-1}}  \\
    &\sim \rho(E) e_q^{- b E}.
\end{eqnarray}

Thus we have derived the distribution function \eref{eq:DistributionFunction}, which we used for our fits, building on the framework of generalized statistical mechanics and superstatistics.

Note that the above equations are only valid for the particular case $\rho(E) \sim E^2$ and $g(\beta)$ being a $\chi^2$ distribution. More generally, one has
\begin{equation}
    P_E(E) \sim \rho (E) \int_0^\infty \frac{g(\beta)}{Z(\beta)}e^{-\beta E} d\beta .
\end{equation}

\section{Applying theory to observation}
We provide a thorough derivation of equation \eref{eq:DifferentialIntensity}, that is $J_E = v(E) P_E$, which relates the distribution function from our superstatistical model with the observed differential flux intensity measured by AMS.

The AMS data \cite{Aguilar2017,Aguilar2018} was published in bins of rigidity $R = p c/ Ze$ with atomic number $Z$, electric charge $e$, momentum $p= | \vec{p} | $, $[R] = [\si{\volt}]$ and the corresponding flux measured in units $[J(R)] = \si{[\meter\tothe{-2} \steradian\tothe{-1} \second\tothe{-1} \giga\volt\tothe{-1}]}$. Instead of rigidity we have chosen to investigate the spectrum in respect to kinetic energy per nucleon. To convert the flux dependence from rigidity $R$ to kinetic energy per nucleon $E$, we need to transform the flux $J_R(R) \to J_E(E)$ such that $J_R(R)\text{d}R = J_E(E) \text{d}E$ is conserved. This is a simple transformation of variables and yields 
\begin{eqnarray}
J_E(E) &= \frac{A}{Ze} \frac{E + u}{\sqrt{E(E + 2u)}} J_R(R),
\end{eqnarray}
with $[J_E(E)] = \si{[\meter\tothe{-2} \steradian\tothe{-1} \second\tothe{-1} \giga\electronvolt\tothe{-1}]}$. For better visibility of the accuracy of our fits, we multiplied the flux with $E^{2.7}$, such that the units for the flux in the presented plots are $[\si{\giga\electronvolt^{1.7} \meter\tothe{-2} \steradian\tothe{-1} \second\tothe{-1}}]$. For the atomic number $A$ we refer to AMS \cite{Aguilar2017, Aguilar2018} who inferred the following average abundance of isotopes $^4\text{He}$, $^{12}\text{C}$, $^{16}\text{O}$, $^{6.5}\text{Li}$, $^{8}\text{Be}$ and $^{10.7}\text{Be}$ among the detected nuclei. The measured flux $J$ represents a differential intensity. Thus it counts the number of particles  with energy $E$ (or rigidity $R$) coming from a unit solid angle that pass through a unit surface per unit of time.

Our superstatistical model builds on a distribution function, denoted as $P$, which counts the spatial density of particles within a given momentum/energy range as
\begin{eqnarray}
    \frac{\text{d}N}{\text{d}^3x} \sim P_E(E) \text{d}E \sim P_p(E) \text{d}^3p.
    \label{eq:conservation}
\end{eqnarray}
Analogously to $P_E(E) \sim \rho(E) e_q^{-b E}$ in the previous section one can derive that $P_p(E) \sim e_q^{-b E}$. 
Thus the density of states $\rho(E)$ can be calculated from the conservation condition \eref{eq:conservation}. Using $E=(\sqrt{p^2+m^2}-m)/A$, which implies that the energy  depends only on the magnitude of the momentum, simplifies $\text{d}^3p = 4 \pi p^2 \text{d}p$, and therefore
\begin{eqnarray}
 P_E(E) \sim \rho(E) e_q^{-b E} \sim  p^2 \frac{\text{d}p}{\text{d}E} e_q^{-b E}.
\end{eqnarray}
Calculating the derivative and using $p^2 = A^2 E(E+2u)$ we find
\begin{eqnarray}
    \rho(E) \sim (E+u) \sqrt{E(E+2u)}.
\label{eq:transf-momentum-energy}
\end{eqnarray}
Note that we generally neglect constant global factors in our equations because we are focusing on the shape of the spectrum rather than its absolute magnitude. Evidently, $[p^2 P_p] = [P_E] = [\si{\electronvolt\tothe{-1} \meter\tothe{-3}}]$ does not have the same dimension as the detected flux, given as differential intensity $J$ with $[J] = [\si{\electronvolt\tothe{-1} \meter\tothe{-2} \second\tothe{-1} \steradian\tothe{-1}}]$ . This reminds us that in order to derive the associated differential intensity from a distribution function we have to account for the rate at which particles go through the detector. That is we multiply with the particle's velocity to obtain the flux $J_E$, corresponding to the distribution function $P_E$, which yields
\begin{eqnarray}
    J_E^{\text{mod}} (E) \sim v(E) P_E(E) \sim v(E) \rho(E) e_q^{-b E}.
\label{eq:Moraal2}
\end{eqnarray}
\cite{Moraal2013} provides a detailed overview about the different ways to count particles including this relation. Evidently, it yields the desired physical dimensions since $[v P_E] = [\si{\electronvolt \meter\tothe{-2} \second\tothe{-1}}]$. 

In order to express the velocity in terms of $E$ we use $p= \gamma m v$ with $\gamma = \frac{1}{\sqrt{1-v}}$ (in $c=1$ convention), $p = A\sqrt{E(E+2u)}$ and $m = A u$ to find
\begin{eqnarray}
    v(E) = \frac{\sqrt{E(E+2u)}}{(E+u)}.
\end{eqnarray}

Plugging everything into \eref{eq:Moraal2} reveals the relation between q-exponential distribution function and the observed differential intensity
\begin{eqnarray}
J_E^{mod} (E)= C  E(E+2u) e_q^{-b E}
\end{eqnarray}
which recovers the function we fitted to the data \eref{eq:FittingFunction}.

\section*{References}

\bibliography{references}

\end{document}